# A Novel Preprocessing Unit for Effective Deep Learning based Classification and Grading of Diabetic Retinopathy

**Pranoti Nage[1*], Sanjay Shitole[2]**

*[1*]Computer Science & Technology, Usha Mittal Institute of Technology, Mumbai, India*
*[2]Information Technology Usha Mittal Institute of Technology Mumbai, India Sanjay.Shitole@umit.sndt.ac.in*

**Abstract—**
Early detection of diabetic retinopathy (DR) is crucial as it allows for timely intervention, preventing vision loss and enabling effective management of diabetic complications. This research performs detection of DR and DME at an early stage through the proposed framework which includes three stages: preprocessing, segmentation, feature extraction, and classification. In the preprocessing stage, noise filtering is performed by fuzzy filtering, artefact removal is performed by non-linear diffusion filtering, and the contrast improvement is performed by a novel filter called Adaptive Variable Distance Speckle (AVDS) filter. The AVDS filter employs four distance calculation methods such as Euclidean, Bhattacharya, Manhattan, and Hamming. The filter adaptively chooses a distance method which produces the highest contrast value amongst all 3 methods. From the analysis, hamming distance method was found to achieve better results for contrast and Euclidean distance showing less error value with high PSNR. The segmentation stage is performed using Improved Mask-Regional Convolutional Neural Networks (Mask RCNN). In the final stage, feature extraction and classification using novel Self-Spatial Attention infused VGG-16 (SSA-VGG-16), which effectively captures both global contextual relationships and critical spatial regions within retinal images, thereby improving the accuracy and robustness of DR and DME detection and grading. The effectiveness of the proposed method is assessed using two distinct datasets: IDRiD and MESSIDOR.

## Introduction

Diabetic retinopathy (DR), a severe microvascular complication induced by Diabetes Mellitus (DM), is one of the major causes of loss of vision worldwide [1, 2]. It is characterized by progressive damage to retinal blood vessels, leading to inadequacy in oxygen supply to retinal tissue [3]. DR has four progressive stages based on severity – mild non-proliferative DR, moderate non-proliferative DR, severe non-proliferative DR and proliferative DR. Diabetic macular edema (DME), caused due to accumulation of fluids and exudates in the macula, is another major complication accelerating vision deterioration. Identification of DR and DME through regular eye examinations is critical for timely treatment administration before the onset of permanent visual impairment [4].

Earlier diagnosis relied on manual evaluation of color fundus photographs by skilled clinicians. However, this approach tended to cause misinterpretations due to inter-observer variability. With advancements in digital fundus imaging and computer-aided screening techniques, automated assessment of DR and DME has received significant attention [5]. Conventional computerized methods employed hand-crafted features based on abnormalities like microaneurysms, hemorrhages and hard exudates. But these features often lacked distinctiveness leading to poor recognition performance.

Automated diagnostic systems for DR have the potential to supersede manual approaches by substantially diminishing the labor-intensive aspects of the screening process. Enhancing efficiency in screening over a broader population is achievable through the system's ability to distinguish between normal and abnormal cases, eliminating the need for manual examination of all images. Hence, the popularity of automatic retinopathy detection systems has surged recently. These systems leverage image processing and computer vision techniques to identify various anomalies associated with retinopathy [6]. As the significance of DR became apparent in the context of diabetes, numerous endeavors were made to precisely categorize its severity and stages. Over time, these classification systems underwent changes, adapting to advancements in understanding the pathophysiology of the disease, evolving imaging techniques for DR assessment, and the emergence of effective treatments. The existing DR classification systems, having proven efficacy, have served as the foundation for substantial research trials and clinical management guidelines for an extended period [7].

Recent research works [8, 9] have investigated Machine Learning (ML) algorithms to extract informative feature representations in an autonomous manner. Deep learning (DL) models such as convolutional neural networks (CNN) have shown impressive improvements in detection accuracy by automatically learning hierarchical inter-relationships from raw images [10]. However, factors like class imbalance, overlapping features and presence of artifacts constrain real-world reliability of existing techniques. The key research problem is enhancing differentiation between normal, DR and DME classes through targeted preprocessing, segmentation and advanced deep learning methodologies.

The various stages of DR and DME are shown in figure 1. Both DR and DME are specifically microvascular complications of diabetes, which result from prolonged high blood sugar levels damaging the small blood vessels in the retina. DR involves the progressive deterioration of retinal blood vessels, leading to various stages of severity, from mild non-proliferative (NP) changes to the advanced stage of proliferative DR (PDR), where abnormal blood vessels grow. DME, a complication of DR, occurs when damaged blood vessels leak fluid into the macula, causing swelling and vision impairment. Both conditions, if left untreated, can lead to significant vision loss.

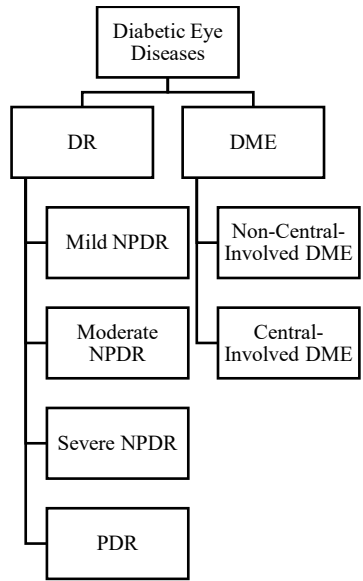

**Figure 1. Various Stages of Diabetic Eye Diseases**

The principal research objectives of this study are –
1. Developing an adaptive preprocessing technique minimizing artifacts and noise while improving image contrast for subsequent analysis
2. Extracting segmented blood vessel regions accurately depicting lesions and abnormalities via improved deep learning algorithms
3. Designing a novel Self-Spatial Attention infused VGG-16 (SSA-VGG-16) based classifier to distinguish between normal, DR and DME classes through learned feature representations
4. Demonstrating consistent performance improvements on standardized datasets compared to current methods

The proposed technique encompasses four phases - preprocessing, blood vessel segmentation, feature extraction and classification.

Fundus preprocessing plays a crucial role in the diagnosis of diabetic-related diseases, particularly DR and DME. These conditions are prevalent among diabetic patients and can lead to severe vision impairment if not detected early. By focusing on these two diagnoses, the preprocessing of fundus images enhances the visibility and clarity of specific retinal features, such as microaneurysms, hemorrhages, and exudates, which are critical indicators of DR and DME. Effective preprocessing, including noise filtering and contrast enhancement, allows for more accurate identification of these subtle yet significant abnormalities. This step not only improves diagnostic accuracy but also supports early intervention and monitoring, making it essential for managing diabetic eye diseases. Hence, this research proposes a novel contrast improvement filter denoted as Adaptive Variable Distance based Speckle (AVDS) filter to enhance the quality of retinal images. The AVDS filter adaptively chooses the distance method (Euclidean, Bhattacharya, Hamming, Manhattan) based on the contrast values obtained.

This facilitates discerning finer details through superior vessel segmentation and ultimately better classification. A customized deep learning architecture called improved Mask Region-based CNN (improved Mask R-CNN) is utilized for precise segmentation of blood vessels. Robust feature extraction is achieved via the SSA-VGG-16 model capturing various abnormalities and patterns. The novel contribution of SSA-VGG-16 lies in its combined use of self-attention and spatial attention mechanisms within the VGG-16 framework, enabling it to capture long-range dependencies and enhance focus on crucial spatial regions. This dual attention approach improves the network's ability to detect subtle retinal abnormalities, leading to more accurate and reliable classification and grading of DR and DME.

Finally, a softmax classifier categorizes retinal images into normal, DR, and DME groups on the basis of extracted feature maps.
The novel contributions of this research are:
1. AVDS filter maximizing contrast for better visualization of retinal components
2. Improved Mask R-CNN segmentation identifying lesions accurately
3. Integrated deep learning framework for enhanced DR & DME recognition
4. Significantly higher sensitivity, specificity and accuracy over existing methods

The remaining sections of the paper are organized as given. Section 2 discusses relevant literature investigating key techniques and findings by earlier studies. Section 3 elaborates the proposed methodology encompassing contrast enhancement, segmentation, feature extraction and classification modules. Section 4 analyzes the experimental outcomes achieved on standardized datasets and provides comparative evaluations with respect to state-of-the-art algorithms. At last, section 5 provides a conclusion to the paper summarizing the critical contributions and suggesting future research directions.

**LITERATURE REVIEW**
Diagnosing DR from fundus images is a labor-intensive task that requires considerable expertise from a professional ophthalmologist. This is particularly challenging in densely populated or remote areas, where both the prevalence of diabetes and DR is expected to surge in the coming years, while the availability of ophthalmologists remains disproportionately low [11-14]. Consequently, the research community has been driven to create computer-aided diagnosis systems to mitigate the challenges, aiming to diminish the cost, time, and effort required by medical experts for DR diagnosis [15]. Over the past decade, computerized screening for DR and DME has been extensively explored to overcome limitations of manual assessment. This section reviews major research dimensions and emerging techniques making the area.

*Conventional/Traditional Approaches:*
Early efforts in DR and DME identification heavily relied on conventional or traditional methods, characterized by manual assessment and dependence on hand-crafted features. The methodology involved in [16] acquired digital retinal images from routine monitoring of DR. A developed automatic analysis tool, comprising statistical classifiers such as Bayesian, Mahalanobis, and KNN, demonstrated superior sensitivity, especially with the Mahalanobis classifier. With potential implications for routine monitoring, this research highlights the value of leveraging digital imaging and automatic statistical analysis systems in improving the efficiency of DR screening and management. The average sensitivity achieved by the model for various abnormalities is 82.75%. In [17], the critical issue of DR was addressed by proposing an automated model for early identification of exudates. Utilizing the template matching algorithm on a dataset of 130 color images of retinal fundus, the system demonstrated impressive results. The emphasis on automated detection showcased the potential for efficient and accurate early diagnosis of DR, aligning with cost-effective healthcare practices and contributing to preventative visual impairment measures. The model shows an accuracy of 98.72% with sensitivity (recall) and specificity rate as 99.45% and 95.68, respectively. The authors of [18] introduced a robust hybrid probabilistic model of learning for DR classification in retinal images. By combining generative and discriminative models, the proposed method utilizes new probabilistic kernels, incorporating Fisher score and information divergences. The hybrid model, featuring a minimum description length criterion, outperforms other methods, demonstrating effectiveness in DR detection with flexibility and valuable applications in data classification. Similarly, a probabilistic learning approach called Gaussian Mixture Model (GMM) was implemented in [19 and 20]. This model represents the probability distribution of the dataset as a mixture of multiple Gaussian distributions. One advantage of GMM is its flexibility in modeling complex and multimodal data, allowing it to adapt well to the diverse characteristics of retinal images.

*Artificial Intelligence (AI) Approaches:*
In response to the constraints of conventional methods, another set of approaches emerged, concentrating on Artificial Intelligence (AI) techniques to enhance the quality of input images. Various organizations have embraced AI, incorporating ML and DL techniques to create automated DR detection algorithms. Some state-of-the-art models are already accessible commercially. These technologies have been crafted using diverse training datasets and varied technical approaches [21]. These models played a pivotal role in improving image quality, thereby laying the foundation for more effective automated diagnosis of DR.

*Machine Learning (ML) Strategies:*

As technology progressed, there was a shift towards machine learning (ML) strategies, signifying a departure from manual and rule-based approaches. A segment-based learning approach was proposed in [22]. The study leverages the benefits of annotating images at a segment level, reducing the burden of expert annotations. By adapting a pre-trained Convolutional Neural Network (CNN) for segment-level DR prediction and integrating all segment levels, the approach achieves a remarkable Area Under the ROC Curve (AUC) of 0.963 on the Kaggle dataset. With a recall and specificity of 96.37%, the end-to-end segment-based approach outperforms existing models, offering improved diagnostic capabilities for DR images. In [23], the datasets encompassing different DR stages were obtained from 500 patients, introducing a novel clustering-based automated region growth architecture. Textures analysis involved extracting features from histograms, wavelets, co-occurrence matrices, and run-length matrices. Various ML classifiers achieved impressive classification accuracies for individual features. To enhance accuracy further, a fused hybrid-feature dataset was created, and optimized features were selected using different techniques. Deploying five ML classifiers on the selected optimized features yielded remarkably high accuracies, reaching up to 99.73%. This research demonstrates the effectiveness of ML techniques in accurate DR segmentation and classification. The authors of [24] classified specialized retinal images using OPF and Restricted Boltzmann Machines (RBM) models. The RBM and OPF approaches, after extracting 500 and 1000 features during training, demonstrated efficacy in recognizing patterns associated with retinopathy and normality. With 15 experiment series and 30 cycles each, the study included 73 diabetic individuals (122 eyes), showing that RBM-1000 achieved the best diagnostic accuracy of about 89.47%. The RBM models exhibited notable sensitivity and specificity in automatic disease detection, particularly in DR screening.

*Deep Learning (DL) Strategies:*
Recent studies underscore the significance of tailored preprocessing in conjunction with deep learning, representing a fusion of techniques to address specific challenges in DR and DME detection. In [25], the limitations of traditional ML algorithms of detecting and classifying DR levels are addressed. Leveraging the power of DL, specifically transfer learning, the study proposes a DL network trained using image features and metadata from a variety of DR fundus images. This approach overcomes the challenges of creating models from scratch and demonstrates the model's capacity to identify unseen fundus images precisely. Trained on IDRD images, the DL model, coupled with different classifiers, exhibits accuracy of upto 95.9%, marking a significant advancement in DR severity classification. The study in [26] employed a systematic approach to analyze fundus scans, involving pre-processing and segmentation techniques, particularly using maximal principal curvature for blood vessel extraction. Post-segmentation, adaptive histogram equalization (AHE) and morphological openings improve and refine the results. A CNN with a unique architecture, incorporating squeeze and excitation, bottleneck layers, and convolutional and pooling layers, was utilized for classification between diabetic and normal retinas. The proposed algorithm, evaluated on DIARETDB1 dataset and a medical institution's dataset, outperforms traditional methods, achieving an accuracy of 98.7% and precision of 97.2%. These results signify the efficiency of the proposed model in identifying DR. The authors of [27] employed a DL network, leveraging AlexNet and ResNet101-based extraction of features, to automatically identify and classify DR fundus images based on severity. Utilizing interconnected layers for identifying critical features and incorporating Ant Colony systems for attribute selection, the chosen characteristics are passed through SVM with multiple kernels, resulting in a final model of classification with excellent accuracy. The approach, relying on 750 features, obtained an accuracy of 93% in DR image classification.

*Local Contrast Enhancement:*
The authors of [30] created a feature map cyclic shift mechanism, where the authors have broken down a traditional local contrast measurement approach into a depthwise parameter-free nonlinear feature refinement layer within a complete network architecture. This layer captures extensive contextual interactions over relatively large distances, providing clear physical insights into the data. Similarly, local contrast based method was employed in [31], where a novel spatial local contrast (SLC) and a novel temporal local contrast (TLC) were combined as STLCF to improve the contrast of the target.

*Scope of Research*
The existing works in DR detection have made commendable progress, transitioning from traditional methods to advanced DL techniques. However, several scopes for improvement exist within this research domain. Firstly, there is room for enhancing the interpretability of deep learning models to make their decision-making processes more transparent and clinically meaningful. Integrating explainable AI techniques could contribute to building trust in the diagnostic outcomes.
In our previous work [28], a DL framework was proposed to detect & classify DR and DME. The framework comprises of preprocessing unit to enhance the image quality by filtering noise, removing artifacts, and enhancing contrast. Blood vessel segmentation was performed via Improved Mask-RCNN. Extraction of features and classification were conducted with SSA-VGG-16 categorizing images into DR, DME, and normal classes, followed by severity level assessment using conditional entropy. Furthermore, the optimization of preprocessing techniques remains a crucial area for refinement. While recent efforts have focused on improving the quality of input images, tailoring preprocessing methods specifically for diabetic retinopathy characteristics could yield more accurate and reliable results. Exploring novel artifact removal algorithms and illumination normalization approaches targeted at the unique features of diabetic retinopathy lesions could enhance the overall efficacy of the detection system.

Additionally, addressing the challenges related to dataset diversity and size is pivotal. As the field evolves, efforts should be directed towards constructing comprehensive and diverse datasets that encompass a broad spectrum of DR manifestations. This will facilitate the development of robust models capable of generalizing well to various clinical scenarios. In summary, the scope of research lies in refining interpretability, optimizing preprocessing techniques, and addressing challenges related to dataset diversity to further advance the capabilities of diabetic retinopathy detection systems.

*Motivation*
The Literature review brings out two crucial aspects which are the underlying motivation for the proposed technique -
1. Importance of preprocessing for reducing noise and improving contrast which aids in discerning retinal lesions accurately
2. Leveraging deep learning methods enhance feature differentiation capabilities for reliable DR and DME recognition
A special variant of speckle filter [29] caught our attention in contrast enhancement, where a novel windowing technique was introduced to divide the total window into five overlapping sub-windows of equal size. Each sub-window contributes to a weighted mean for the pixel being filtered, with weights determined by sub-window heterogeneity measures. The filter employs Quadratic Corner Difference (QCD) to create masks for the windowing method of speckle filtering. This filter automatically adjusts speckle suppression strength based on local statistics, ensuring edge preservation while effectively reducing speckle in homogeneous areas.

Motivated by such local contrast enhancement based method, the proposed model employs a novel AVDS filter to minimize artifacts and boost image quality which allows finer segmentation of abnormal regions. However, in the proposed AVDS, four distance methods are replaced with QCD, they are: Euclidean, Bhattacharya, Manhattan, and Hamming. The best method out of the 4 is chosen and proceeded with the rest of the framework, where, an improved Mask R-CNN facilitates precise localization of manifestations through robust encoder-decoder architecture. And, the extracted feature maps are classified using a customized SSA-VGG-16 model ensuring high sensitivity and specificity. The integrated methodology is anticipated to push performance boundaries over current approaches.

**PROPOSED METHODOLOGY**
The overall architecture of the proposed technique consists of four main units as illustrated in Figure 2. First, fundus images are enhanced using fuzzy filtering for the removal of noise, non-linear diffusion filtering for the removal of artefacts, and the newly developed AVDS filter for contrast improvement. Next, segmentation of blood vessels is conducted via improved Mask R-CNN focusing especially on abnormal regions. The segmented image is then passed to a SSA-VGG-16 model which performs multi-level feature extraction encoding various lesions and patterns. Finally, a softmax classifier categorizes the fundus image as normal, DR or DME based on the learned features. The upcoming subsections explain the individual components of proposed model in a detailed manner.

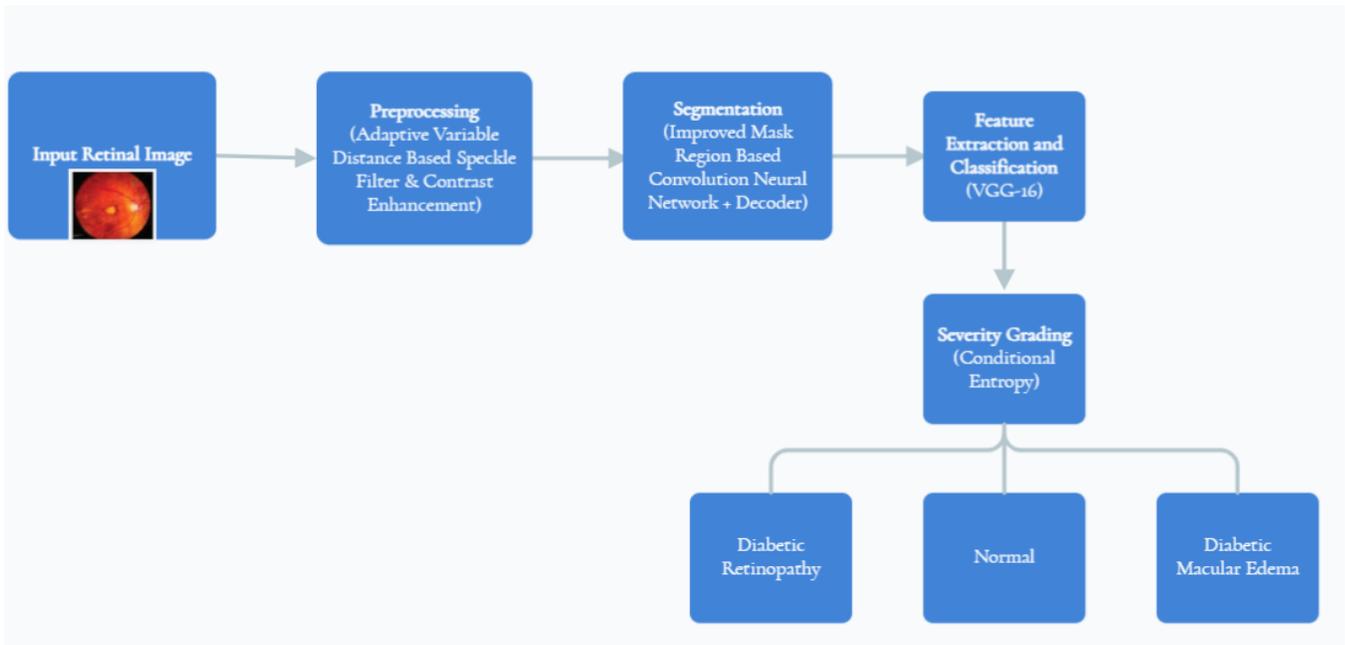

**Figure 2. Proposed Framework**

*Preprocessing Stage*
This represents the preliminary phase in the detection process for DR and DME. The preprocessing unit consists of fuzzy filter for noise removal, non-linear diffusion filter for artifacts removal, and proposed novel AVDS for contrast enhancement. The pseudocode of the preprocessing technique is given below and the explanation of each of the blocks of preprocessing unit is given in the following sub-sections.

> **Algorithm 1: Preprocessing Stage**
> 1) Apply Novel AVDS Filter for Contrast Enhancement:
> Function NovelAVDSFilter(image):

```
For each pixel (x, y) in the image:
Compute Euclidean Distance (ED) using Equation (3)
Compute Hamming Distance (HD) using Equation (4)
Compute Bhattacharya Distance (BD) using Equation (5)
Compute Manhattan Distance (MD) using Equation (6)
Determine the distance yielding maximum contrast
End For
Apply the chosen distance function to enhance contrast
Return enhanced image
```

### Noise Removal using Fuzzy Filter

Fundus images often encounter challenges due to the presence of both Gaussian and impulsive noises. Existing methods, such as statistical filters, have proven successful in effectively removing Gaussian noise, and impulsive noise has been addressed using median filters with success (BahadarKhan et al., 2016).

On the other hand, fuzzy filtering is particularly well-suited for noise filtering in medical image processing due to its ability to handle the inherent uncertainty and variability in medical images. Medical images often contain complex, nuanced details that are critical for accurate diagnosis and analysis. Fuzzy filtering allows for precise noise reduction while preserving essential image features, such as edges and textures, which are vital in distinguishing between healthy and pathological areas. Its adaptive nature is beneficial for medical imaging, as it can effectively manage variations in image quality without compromising critical diagnostic information. Therefore, despite the existence of more advanced methods, fuzzy filtering remains a valuable tool in medical image processing where maintaining data integrity and detail preservation is paramount. However, when faced with the task of removing mixed noise types from retinal images, these conventional approaches struggle and often result in undesired blurriness. To overcome this limitation, we introduce a fuzzy filtering approach for noise removal, specifically designed to handle the complexity of mixed noise in retinal images.

In our proposed method, we acknowledge the difficulties posed by mixed noise types and address them through the incorporation of a fuzzy filter. This innovative approach aims to remove both Gaussian and impulsive noises seamlessly, providing a solution to the challenges faced by conventional methods. The key concept driving the success of our fuzzy filter lies in its ability to manage uncertainty through the utilization of membership function variables.

Let 'm' represent the corrupted image affected by noise, and consider a pixel positioned at (x, y). The noise removal process is defined mathematically as follows:

$$I_{x,y} = \frac{\sum_{i=-n}^{n}\sum_{j=-n}^{n} W(\Delta Pi,j)Pi,j}{\sum_{i=-n}^{n}\sum_{j=-n}^{n} W(\Delta Pi,j)} \quad (1)$$

Where $\Delta Pi, j = Pi, j - Pm, n$ denotes the grey level difference between the pixel in the center and its neighboring pixels on all 4 sides, with 'n' representing the pixel size. The weight values $W(\Delta Pi, j)$ are computed on the basis of grey level difference of the neighboring pixels. The term $W(\Delta Pi, j)$ is expressed as the membership function of crisp values, where a crisp value of one indicates a low grey level difference, and a crisp value of zero indicates a high grey level difference. $Pm, n$ represents the intensity value of the central pixel located at coordinates (m, n).

To elaborate, the membership function plays a vital part in the identification of weight assigned to each pixel with respect to its grey level difference. The membership function ensures that pixels with low grey level differences are given more weight, signifying their importance in the filtering process. Conversely, pixels with high grey level differences receive lower weights, reflecting their lesser impact on noise removal.

Additionally, a dynamic threshold T is calculated, taking into account the resolution of the image. This dynamic threshold is crucial for adapting the noise removal process to the specific characteristics of each image, considering that the resolution of each image varies. The dynamic threshold adjustment is an integral part of the fuzzy filtering approach, contributing to its effectiveness in noise removal.

### Artefacts Removal using Non-Linear Diffusion Filter

The Nonlinear Diffusion Filtering technique is utilized to address artifacts, emphasizing the correction of blurriness, illumination challenges, and poor edges, thereby indirectly improving image quality based on edge preservation and illumination correction. The algorithm operates within the image domain μ with $F_a()$ representing the original image. The resulting filtered image $x_{a,t}()$ is obtained through a nonlinear diffusion function with the initial state set as the original image. The algorithm involves partial differential equations (PDEs) that govern the evolution of the image over time (t), ensuring the preservation of important image features. The diffusivity variable $g$ is strategically chosen to act as an edge detector, minimizing edge smoothing during the diffusion process. This approach effectively eliminates blurriness, poor edges, and illumination issues from the images while preserving essential image details. The algorithm contributes to enhancing image quality and preparing the images for subsequent processing steps in the overall technique.

$$\partial_t X = \text{div}(g(|\nabla x_\alpha|^2)\nabla x) \text{ on } \mu \times (0, \infty)$$
$$X(a, 0) = F(a) \text{ on } \mu$$
$$\partial_n X = 0 \text{ on } \partial\infty \text{ x } (0, \infty)$$

$$g(v^2) := \begin{cases} 1, & (v^2 = 0) \\ 1 - \exp\left(-c \Big/ \left(\frac{v}{\lambda}\right)^8\right), & (v^2 > 0) \end{cases} \quad (2)$$

***Contrast Enhancement using Novel AVDS Filter***
A filtering operation generally involves averaging the pixels within a window that is symmetrically distributed around a central pixel. This process can cause high-contrast areas or edges present in one part of the central pixel's neighborhood to spread to other parts where they were not initially present, resulting in edge blurring. To address this issue, our work employs a subwindow-based approach. This method prevents high-contrast areas or edges in one part of the neighborhood from affecting other parts.

The AVDS filter represents an adaptive technique for contrast improvement in retinal images through novel distance computation. Existing variant of speckle filtering methods employs QCD method to calculate pixel statistics within local regions and suppress noise accordingly [29]. In fundus image analysis, several types of noise can significantly affect image quality and hinder accurate diagnosis. These include Gaussian noise, which is commonly introduced during image acquisition, and impulsive noise, which often arises from transmission errors or sensor faults. Gaussian noise can blur subtle retinal features, while impulsive noise may result in random bright or dark spots that obscure critical details. Additionally, fundus images may suffer from uneven illumination, causing varying brightness levels across the image, and motion artifacts, which can lead to blurring and distortions.

The proposed AVDS technique addresses these issues by adaptively selecting the optimal distance metric for each image region, which enhances contrast while preserving essential details. Unlike traditional methods that might use fixed thresholding, AVDS dynamically responds to the specific characteristics of the image, allowing it to handle both Gaussian and impulsive noise effectively. Furthermore, the AVDS filter is designed to minimize over-smoothing, which is a common issue with traditional filtering techniques that can obscure important retinal features. This adaptive approach ensures that the technique is versatile and capable of improving image quality across a range of noise conditions, making it particularly suitable for the challenges associated with fundus image analysis.

However, this often leads to over-smoothing causing loss of finer details critical for clinical interpretation. The newly introduced AVDS filter computes four different distances to capture local contrasts, they are - Euclidean, Hamming, Bhattacharya and Manhattan distances. The proposed AVDS technique enhances both high- and low-contrast fundus images by dynamically selecting the most suitable distance metric (Euclidean, Hamming, Bhattacharya, or Manhattan) based on the local contrast within each region of the image. This adaptive selection ensures that the AVDS filter maximizes contrast enhancement without requiring fixed thresholds, which allows the technique to effectively handle a wide range of contrast conditions.

Initially, the total filter mask is divided into five sub-windows, each including the pixel to be filtered (central pixel) as one of its elements. This neighborhood structure makes the filter more adaptive to single targets, edges, and homogeneous regions. Generally, if $k$ is the size of each sub-window, then the total filter mask will be of size $2k - 1$. Then the following distances are calculated to choose the pixel to be filtered.

Euclidean Distance: The distance defined by the length of the line connecting the two pixel coordinates.
$$d = \sqrt{(x_2 - x_1)^2 + (y_2 - y_1)^2} \quad (3)$$

Where the coordinates (x1,y1) and (x2,y2) represent the positions of two pixels in the image grid. This distance measures the straight-line intensity difference between these two pixels, which is important for detecting changes in the image that indicate abnormalities such as microaneurysms or hemorrhages.

Hamming Distance: The distance defined by the measure of the difference between two strings of pixels of equal length.
$$HD = \text{sum}(\ xi \neq yi\ ), \text{ where } i = 1, 2 \quad (4)$$

Bhattacharya Distance: The distance defined by the measure of the similarity between two probability distribution functions (PDFs) associated with the coordinates.
$$BD = -\ln\left(\sum \left(\sqrt{p(x_i) \cdot p(y_i)}\right)\right) \quad (5)$$

Where p(.) refers to the PDF of the corresponding coordinate.
Manhattan Distance: The distance defined by the sum of the absolute differences between the coordinates.
$$MD = |\ x1 - y1\ | + |\ x2 - y2\ | \quad (6)$$

Here, x and y are intensity vectors of the reference pixel and its neighbors within the filtering window respectively. The contrast measure is computed for the images resulting from each distance function. The final filtered value I is attained for a mask of size $2k - 1$ as expressed below.

$$I = \frac{\sum_{i=1}^{k} \mu_i \left(\frac{1}{D_i}\right)^\omega}{\sum_{i=1}^{k} \left(\frac{1}{D_i}\right)^\omega} \quad (7)$$

Where $D_i \in \{ED_i, HD_i, BD_i, MD_i\}$, $\mu_i$ is the mean of the i-th sub-window. The sub-window configuration, combined with the inverse D, helps preserve features while the mean operation handles the smoothing. The exponent $\omega$ controls how well the filter adapts to the specific window being analyzed. By adjusting $\omega$, the filter can control the balance between feature preservation and smoothing. A higher value of $\omega$ gives more weight to the inverse D, enhancing the filter's ability to preserve edges and features. Conversely, a lower value of $\omega$ will result in more smoothing, as the mean operation becomes more dominant.

The distance yielding the maximum contrast is adaptively chosen for further processing. By avoiding over-smoothing, the AVDS filter enables finer details of lesions to become more discernible.

***Segmentation Stage***

Improved Mask R-CNN: This enhances the segmentation performance of the popular Mask R-CNN [23] technique through modifications like boundary refinement modules, upsampling decoders and multi-level feature aggregation networks. This facilitates accurate capture of abnormalities within the vasculature structure as shown in figure 3.

After the effective preprocessing steps, the segmentation of blood vessels becomes a critical step in retinal image analysis. This is achieved through the utilization of an advanced algorithm, an upgraded version of the Mask R-CNN. Blood vessels play a pivotal role in calculating image intensity, edges, texture, and various features like the blood vessels, lesions, and optic disc. Analyzing the diagnosis over the segmented regions significantly boosts the accuracy and precision of disease identification.

The selection of Mask R-CNN is based on its simplicity in training compared to existing CNN algorithms, with a specific focus on particular regions, contributing to enhanced efficiency. Nevertheless, the conventional Mask R-CNN has a drawback of low boundary precision accuracy. To address this limitation, our proposed approach employs an improved Mask R-CNN to achieve higher precision accuracy.

The preprocessing involves removing the optic disc from the preprocessed image, and subsequently, the blood vessels are segmented with the help of improved Mask R-CNN. The initial step in this process is the Region of Interest (ROI) alignment, predicting the region of interest from the input image.

### Network Structure of Enhanced Mask R-CNN

The enhanced Mask R-CNN is engineered to concurrently conduct identification of blood vessels and pixel-level blood vessels. It incorporates structures from the faster R-CNN network and the Feature Pyramid Network (FPN) with a Region of Interest (ROI) Align algorithm. The comprehensive structure of the Mask R-CNN is dissected into six blocks: input, backbone network, FPN, Regional Candidate Network (RPN), ROI alignment and bounding box, category, and the mask output (box, class, and mask).

### ROI Alignment

ROI alignment proves to be a crucial step in Mask R-CNN to enhance pixel-level segmentation of blood vessels. The layer of ROI alignment refrains from quantization for the boundary of ROI. It involves achieving information about the optic disc surrounding regions from the retinal fundus images through binary classification in both background and foreground blocks. The improved Mask R-CNN introduces a decoder layer with learnable up-sampling to handle features with high spatial resolution. The features of high-resolution achieved from the ROI are used and aligned to the decoding layer with the help of skip connections.

$$B_j = \{P_1, P_2, P_3, P_4\} \quad (7)$$

Where $B_j$ is described by the base points as given below:

$$ROI(M,N) \approx \frac{ROI(P_1)}{(M_2 - M_1)(N_2 - N_1)} * (M_2 - N)(N_2 - N) +$$
$$\frac{ROI(P_2)}{(M_2 - M_1)(N_2 - N_1)} * (M_2 - M)(N_2 - N) +$$
$$\frac{ROI(P_3)}{(M_2 - M_1)(N_2 - N_1)} * (M_2 - M)(N_2 - N) +$$
$$\frac{ROI(P_2)}{(M_2 - M_1)(N_2 - N_1)} * (M_2 - M)(M_2 - C) \quad (8)$$

### Loss Function

The loss function in Mask R-CNN for all sampled Regions of Interest (ROI) integrates the accumulation of Mask loss, Bounding-box loss, and Classification loss. The loss is a combination of prediction loss for class labels, refinement loss of bounding boxes, and prediction loss for mask segmentation. The loss function ensures a comprehensive assessment for accurate blood vessel segmentation.

$$L' = I'_{cls} + I'_{box} + I'_{mask} \quad (9)$$

$$I.(\Pr_b, G_b, \Pr_{cls}, G_{cls}) = I_{.c}(P_{cls}, G_{cls}) + \varphi[G_{cls} \geq 1] \quad (10)$$

---

**Algorithm 2: Segmentation using Improved Mask R-CNN**
1) Preprocessing:
Remove optic disc from preprocessed image
2) Network Structure of Enhanced Mask R-CNN:
Input:
Input image
Backbone Network:
Extract features from input image
Feature Pyramid Network (FPN):
Generate multi-level feature maps
Regional Candidate Network (RPN):

---

```
Propose candidate object regions
ROI Alignment and Bounding Box:
Refine region proposals and predict bounding boxes
Category and Output of the Mask:
Classify objects and output segmentation masks
3)   ROI Alignment:
Function ROIAlignment(image):
Obtain ROI from input image
Refrain from quantization for ROI boundary
Introduce decoder layer with learnable up-sampling
Align high-resolution features from ROI to decoder layer
Use skip connections to align features
Return aligned features
4)   Loss Function:
Function MultiTaskLoss(P_cls, G_cls, Pr_b, G_b, Pr_cls, G_cls, Pr_mask, G_mask):
Calculate classification loss I_cls
Calculate bounding box loss I_box
Calculate mask loss I_mask
Combine losses to get total loss L'
Return total loss L'
```

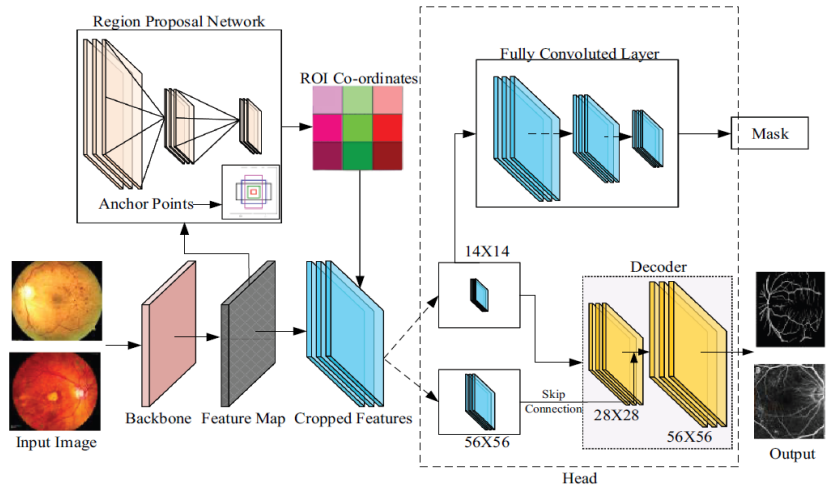

**Figure 3. Process of segmentation of blood vessels using improved mask R-CNN [28]**

*Classification & Grading*
VGG-16 Classifier: This is an established CNN architecture known for its capability to learn robust feature representations facilitated by increased depth [24]. In the proposed method, VGG-16 extracts numerous lesions, texture and morphological attributes which are classified using additional layers.

After completing the segmentation phase, the segmented images undergo feature extraction and classification. The proposed SSA-VGG-16 architecture is applied for this task. SSA-VGG-16 comprises convolutional layers, fully connected layers, and softmax layers. The proposed SSA-VGG-16 model starts with the foundational VGG-16 architecture, which consists of a series of convolutional layers followed by pooling layers, and culminates with fully connected layers for classification. The model is modified to include self-attention and spatial attention mechanisms, collectively referred to as SSA, which are added after the last convolutional block of VGG-16. This attention module ensures that the network can capture global context as well as focus on crucial spatial details within the retinal images.

In the SSA module, a self-attention block is first applied to capture long-range dependencies between spatial locations in the feature map. This mechanism enables each pixel to attend to all other pixels, creating an attention map that represents how each location in the image is influenced by all other locations. The self-attention mechanism helps the model to focus on critical regions of the retina, ensuring that subtle features related to retinal damage are detected. Meanwhile, the spatial attention mechanism refines this focus by highlighting important regions in the feature maps, such as areas with swelling or fluid accumulation, which are characteristic of DME. The feature map from the last convolutional block is transformed into query, key, and value matrices. These matrices are used to compute attention weights, representing the similarity between each pair of pixels. Using dot-product attention, the similarity between the query and key matrices is calculated, scaled, and passed through a softmax function to generate attention scores. These scores are then used to weight the value matrix, allowing the model to aggregate information from all locations in the feature map. This self-attention process results in a context-aware

*A Novel Preprocessing Unit for Effective Deep Learning based Classification and Grading of Diabetic Retinopathy*

feature representation that enhances the model's sensitivity to relevant features across the entire image.
Following the self-attention block, a spatial attention mechanism is applied to further refine the focus on key regions of interest within the feature map. Average and max pooling operations are performed along the channel dimension to summarize the presence of important features at each spatial location. The pooled features are concatenated and passed through a convolution layer with a sigmoid activation function, generating a spatial attention map that highlights the significant regions. This attention map is then used to scale the original feature map, thereby emphasizing spatial areas that are crucial for identifying and grading DR and DME. By combining self-attention and spatial attention, the SSA module ensures that the model is both contextually aware and spatially focused, making it particularly adept at identifying subtle retinal abnormalities associated with DR and DME.
Structural features like microaneurysms, hemorrhages, and hard exudates are focused on, alongside shape, orientation, and color features, contributing to the comprehensive analysis for the classification of DR and DME.

$$C(F) = \vartheta(f^{7\times 7}[\,F_a; F_m\,]) \quad (11)$$

In feature extraction, both max pooling and average pooling layers contribute to extracting relevant features from the segmented image. The resulting output, denoted as C(F), is a critical step in the feature extraction process.

$$C\left(p = \frac{S}{t}\right) = e^d / \Sigma_j e^{d_j} \quad (12)$$

The extracted features move on to a fully connected layer, incorporating a dense layer, dropout layer, and flatten layer. The softmax layer is pivotal in classifying these features into three distinct categories: normal, DR, and DME. The classification process involves calculating weight values, and the output is categorized into the three classes using a probability formula.

$$H(X) = -\sum_{x \in X} P(X) \log(X) \quad (13)$$

The entire process of extraction of features and classification is depicted in Figure 4. Following this, the images are categorized as normal, DR, or DME. To assess the disease's severity, an entropy function is employed, factoring in the total count of lesions for threshold generation. Severity levels, ranging from mild to moderate and severe, are assigned based on the computed threshold value from the entropy function.

**Algorithm 3: Feature Extraction and Classification**
Initialise Features F = {f1, f2, ..., fn}
Initialise SSA-VGG-16 model
SSA-VGG-16.train(training_data)
For i = 0 to n do
Extract F from segmented region
Feature extracted by Fa
Feature extracted by Fm
Combined_features = Combine(Fa, Fm)
Classify_result = SSA-VGG-16.classify(Combined_features)
Assign class labels: Class = {normal, DR, DME}
Output Classify_result
End for

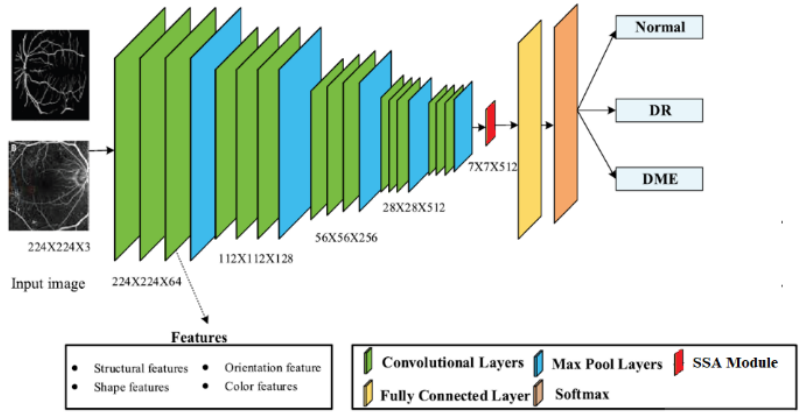

**Figure 4. Process of extraction of features and classification using SSA-VGG-16 [28]**

**RESULTS & DISCUSSION**
*Dataset Description*
*IDRiD Dataset*
The IDRiD (Indian Diabetic Retinopathy Image Dataset) is an innovative collection designed specifically for the Indian demographic, capturing both the details of diabetic retinopathy lesions and the nuances of normal retinal structures down to the pixel. This rich dataset is a game-changer for refining image analysis techniques aimed at catching diabetic retinopathy in its early stages. It's neatly organized into three main parts:

Segmentation, Disease Grading, and Localization. In the Segmentation area, you'll find original color images of the retina alongside detailed annotations for lesions and the optic disc. The Disease Grading section includes these retinal images but focuses on identifying the severity levels of diabetic retinopathy and diabetic macular edema. Meanwhile, the Localization segment highlights the exact positions of key retinal features like the optic disc and fovea center. IDRiD is more than just a dataset; it's a critical asset for those on the front lines of creating and testing new tools to combat diabetic eye conditions, offering a clear window into both healthy and diseased states.

*Messidor Dataset*
The MESSIDOR dataset, short for Methods to Evaluate Segmentation and Indexing Techniques in Retinal Ophthalmology, features a collection of 1200 color images of the eye's fundus, focusing on the posterior pole. These images were captured by three different ophthalmologic departments using a sophisticated color video 3CCD camera attached to a Topcon TRC NW6 non-mydriatic retinograph, which provides a broad 45-degree view of the retina. The images are detailed, captured at 8 bits per color plane, and come in various resolutions (1440*960, 2240*1488, or 2304*1536 pixels). Of these, 800 images were taken with the pupils dilated, and 400 were taken without dilation. Organized into three sets to represent each ophthalmologic department, the dataset is further segmented into four zipped subsets containing 100 TIFF format images each. To aid in medical analysis, each subset is accompanied by an Excel file detailing the medical diagnoses for every image, including the grade of retinopathy and the risk of macular edema present. This carefully curated dataset is an invaluable tool for those looking to advance segmentation and indexing methods in the field of retinal ophthalmology.

*Results of Preprocessing Unit*
Performance measures considered for evaluating the preprocessing unit are MSE, RMSE, PSNR, & Contrast.

*MSE*
Mean Squared Error (MSE) is a performance metric used for evaluating the performance of a preprocessor or a regression model by measuring the average squared difference between the predicted and actual values. It quantifies the average squared deviation of predictions from the ground truth.
MSE calculates the average of the squared differences between actual and predicted values, providing a measure of how well the model or preprocessor is performing. Lower MSE values indicate better performance, with zero MSE representing a perfect match between actual and predicted values. The formula for MSE is as follows:

$$MSE = (1/n) * \Sigma(y_i - \hat{y}_i)^2 \quad (14)$$

Where n denotes the number of samples in the dataset; $y_i$ represents the actual value of the i-th pixel; $\hat{y}_i$ represents the predicted value of the i-th pixel; and $\Sigma$ denotes the summation across all pixels of the image.

*RMSE*
Root Mean Squared Error (RMSE) is a variation of the MSE commonly used for evaluating the performance of regression models or preprocessors. RMSE is advantageous because it presents the error metric in the same unit as the target variable, making it easier to interpret.
RMSE evaluates the square root of the average of the squared differences between predicted and actual values. This metric renders a more intuitive understanding of the error by bringing it back to the original unit of the target variable. As with MSE, lower RMSE values indicate better model or preprocessor performance.

$$RMSE = \text{sqrt}(MSE) \quad (15)$$

*PSNR*
Peak Signal-to-Noise Ratio (PSNR) is a performance measure used for computing the quality of a processed or compressed signal concerning the original signal. It provides insight into the level of distortion introduced during compression or processing. PSNR is expressed in decibels (dB), and a higher value of denotes a smaller amount of signal distortion or noise. It is a widely utilized measure in image and video processing to assess the visual quality of compressed or processed signals.

$$PSNR = 20*\log_{10}(MAX) - 10*\log_{10}(MSE) \quad (16)$$

*Contrast*
Contrast in an image refers to the difference in intensity between the darkest and lightest parts of the image. In the context of a preprocessed image, contrast enhancement techniques are often applied to improve the visibility of details by increasing the difference in intensity between different regions.
Calculating contrast value involves measuring the standard deviation of pixel intensities in the image. This contrast value provides a measure of how spread out pixel intensities are in the image. Higher contrast values indicate a more distinct difference between light and dark areas, leading to a visually sharper image. Contrast enhancement is a common preprocessing step to improve the quality and visibility of important features in images.

*Results of Classification & Grading Unit*
Performance measures considered are accuracy, sensitivity, specificity, f1-score, and ROC curve.

*Accuracy*
Accuracy, in the context of a confusion matrix, is a key indicator of how well a classification model performs, quantifying the proportion of predictions it got right. It takes into account both the true positives (correctly identified instances) and the true negatives (correctly rejected instances), comparing these to the overall count of instances being examined. The formula to compute accuracy is:

$$Accuracy = (TP + TN) / (TP + TN + FP + FN) \quad (17)$$

Where:
- TP: True Positives
- TN: True Negatives
- FP: False Positives
- FN: False Negatives

Accuracy calculates the proportion of instances that are correctly predicted (including both true positives and true negatives) relative to the entire dataset's number of instances.

At its core, accuracy evaluates how well the model can correctly identify both positive and negative instances. However, in situations where there's an uneven distribution of classes, accuracy may not be the most reliable measure. This is because a model might achieve a high accuracy score by predominantly predicting the more frequently occurring class, skewing the real picture of its performance. For a more comprehensive evaluation, additional metrics like precision, recall, and F1 score may be considered alongside accuracy. Accuracy of the proposed work is 98.7% for IDRiD dataset and 98.2% for Messidor dataset.

*Sensitivity*
Sensitivity, also called as True Positive Rate (TPR) or Recall, is a performance metric from a confusion matrix that measures the capability of a classification algorithm to correctly identify positive cases from the entire pool of positive cases.
Sensitivity = (TP) / (TP + FN)     (18)

In the medical context, sensitivity is crucial, as it indicates the model's effectiveness in capturing instances of a particular condition. A high sensitivity value implies that the algorithm has a low rate of FN, ensuring that most actual positive cases are correctly identified. However, there is often a trade-off between sensitivity and specificity, and the appropriate balance depends on the specific goals and constraints of the classification task. Sensitivity of the proposed work is 98.2% for IDRiD dataset and 98.5% for Messidor dataset.

*Specificity*
Specificity is a performance metric derived from a confusion matrix in the context of classification models. It gauges the model's capability to accurately distinguish negative instances among all actual negative cases.
Specificity = (TN) / (TN + FP)     (19)

Specificity is essential in situations where correctly identifying true negatives is crucial, such as in medical diagnostics or other scenarios where false positives can have significant consequences. A high specificity value denotes that the model has a low rate of FPs, meaning that it accurately identifies instances that are truly negative. Similar to sensitivity, there's often a trade-off between sensitivity and specificity, and the optimal balance depends on the specific needs of the classification problem. Specificity of the proposed work is 99.1% for IDRiD dataset and 98.9% for Messidor dataset.

*AUC*
The Area Under the Curve (AUC) is the area under the Receiver Operating Characteristic (ROC), which is a plot showing the trade-off between true positive rate (sensitivity) and false positive rate (FPR) (1 - specificity).
AUC = ∫[TPR (FPR)] d(FPR)     (20)

The AUC of the proposed model is 0.95 for IDRiD dataset and 0.94 for Messidor dataset.

*Comparative Evaluation*
The study compares existing preprocessing methods and the proposed model of preprocessing with respect to MSE, RMSE, PSNR, & Contrast measure. The outcomes are tabulated in Table 1.

**Table 1. Performance Comparison of Preprocessing Methods**

| Method | MSE | RMSE | PSNR | Contrast |
|---|---|---|---|---|
| **Histogram Equalization** | 105.77 | 10.28 | 26.89 | 10.23 |
| **CLAHE** | 109.84 | 10.48 | 27.11 | 11.56 |
| **Proposed Euclidean based AVDS Method** | 80.04 | 8.94 | 29.09 | 12.31 |
| **Proposed Bhattacharya based AVDS Method** | 85.56 | 9.24 | 28.8 | 18.18 |
| **Proposed Hamming based AVDS Method** | 93.48 | 9.66 | 28.42 | 50.23 |
| **Proposed Manhattan based AVDS Method** | 79.76 | 8.93 | 29.09 | 12.26 |

Thus the table 1 proves that the performance of the proposed AVDS method of filtering involves four distance methods such as Euclidean, Bhattacharya, Hamming, and Manhattan, outperform the existing methods Histogram Equalization and CLAHE by exhibiting least values for MSE and RMSE and highest values of PSNR and contrast. Euclidean based approach is found to perform the best in lower MSE and higher PSNR values with hamming distance showing best contrast value and so, the further processing is carried out using the Euclidean and hamming based AVDS output.

Furthermore, table 1 justifies the effectiveness of the novel AVDS filter in improving image quality and minimizing artifacts. The lower MSE and RMSE values, particularly for the Euclidean and Manhattan distance-based AVDS methods, indicate superior noise reduction, while the higher PSNR values demonstrate better signal preservation compared to traditional methods like Histogram Equalization and CLAHE. Additionally, the significantly improved contrast, especially with the Hamming distance-based AVDS filter, enhances the visibility of critical retinal features, which is essential for accurate diagnosis of DR and DME. These metrics collectively validate the AVDS filter's ability to boost image quality and handle noise and artifacts more effectively than conventional techniques.

The image results taken in BGR format of the 4 variants of AVDS are illustrated along with histograms in figure 5.

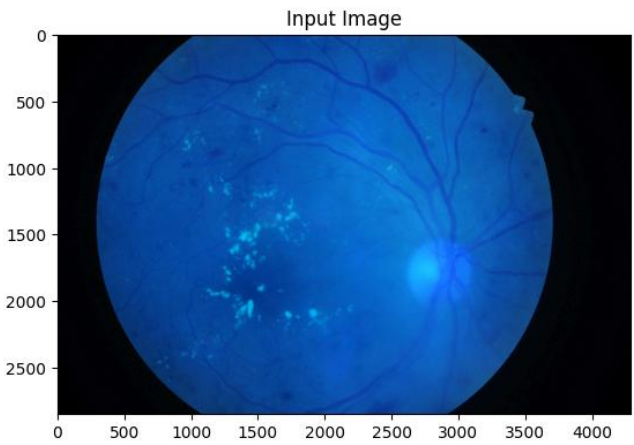
**(a) Input Image**

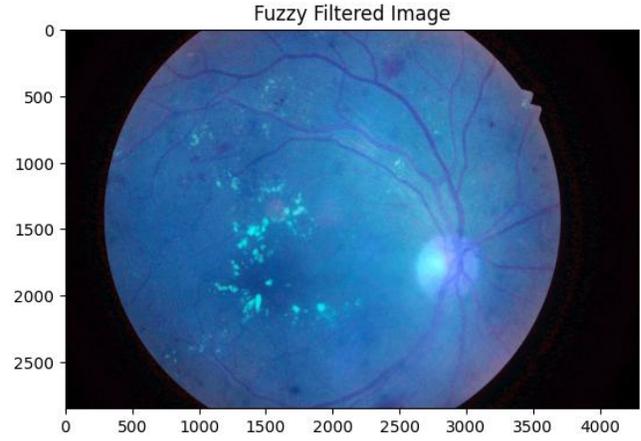
**(b) Fuzzy Filtered Image**

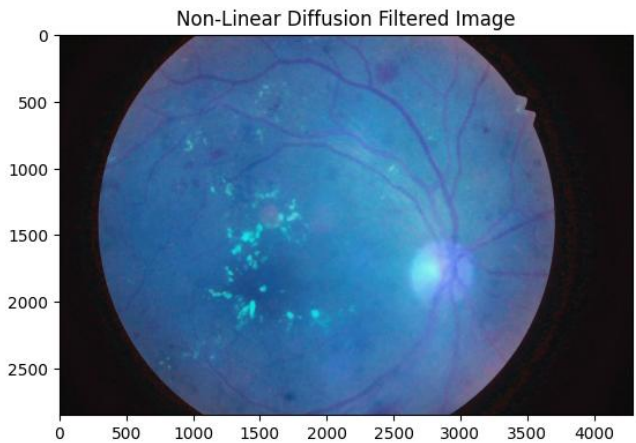
**(c) Non-Diffusion Filtered Image**

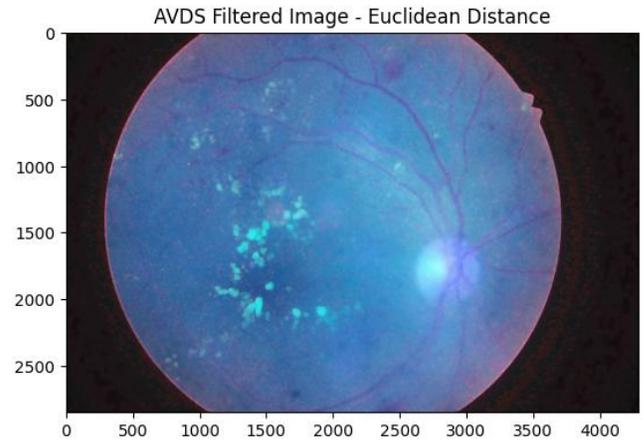
**(d) ADVS – Euclidean Distance Output**

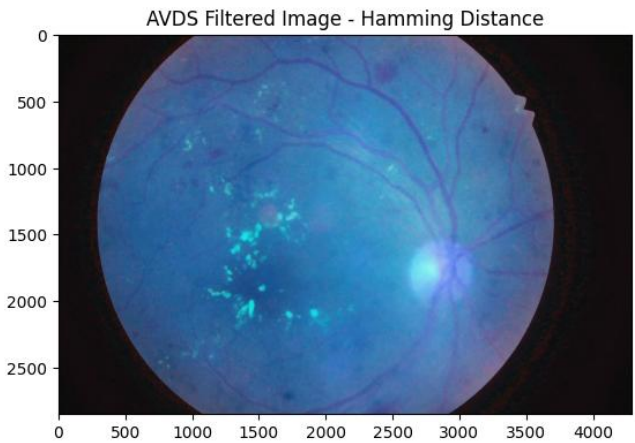
**(e) ADVS – Hamming Distance Output**

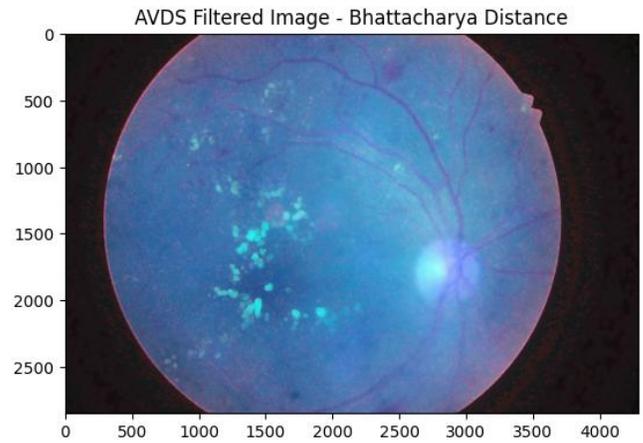
**(f) ADVS – Bhattacharya Distance Output**

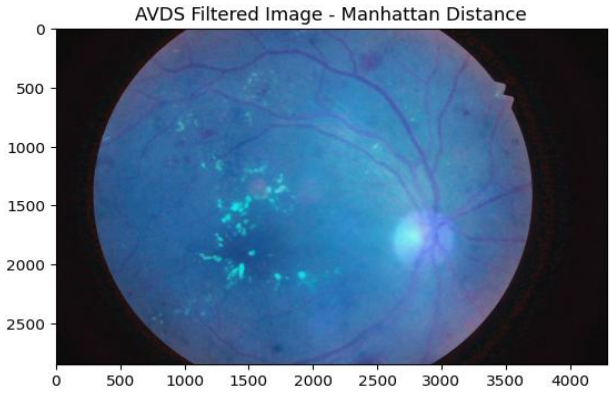 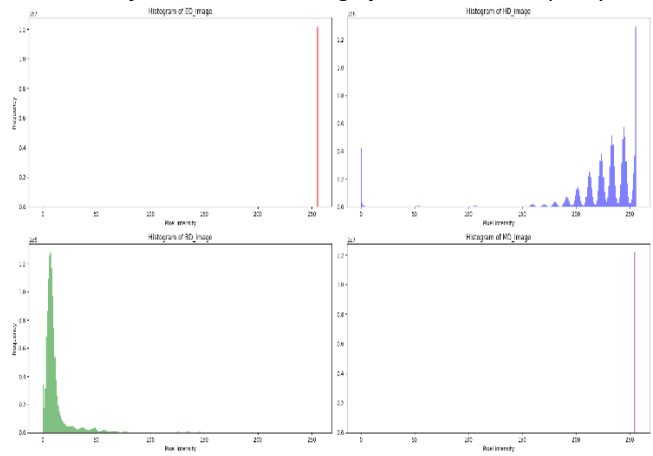

**(g) ADVS – Manhattan Distance Output**  **(h) Histogram Plots of the 4 variants of AVDS**

**Figure 5. AVDS Filtered Image Results based on Distance Metrics**

From the histograms of the processed images using different AVDS metrics, we observed distinct variations in pixel intensity distributions. The AVDS Euclidean output showed an overwhelming concentration of high-intensity pixels. The AVDS Hamming output showed diverse intensity peaks, suggesting a preservation of image details across a broader range of intensities, which could be advantageous for applications requiring detailed textural information. The AVDS Bhattacharya output predominantly displayed low intensities, showing it can be potentially useful for enhancing visibility in darker regions. The AVDS Manhattan output exhibited extreme brightness. Each method has its own characteristics. The choice of which technique is performing well depend heavily on the specific application. In the context of contrast enhancement, AVDS Hamming distance output is found to give best results.

The study compares existing model [28], [32], [33], [34] and the proposed model of classification and grading in terms of accuracies, sensitivities, and specificities.

**Table 2. Performance Comparison of Classification Models**

| Method | Accuracy (%) | Sensitivity (%) | Specificity (%) | AUC |
|---|---|---|---|---|
| **Ref [28]** | 80.7 | 93.67 | 93.67 | 0.93 |
| **Ref [32]** | 90.07 | - | - | - |
| **Ref [33]** | 95.65 | 89 | 99 | - |
| **Ref [34]** | 94.17 | 94.17 | - | - |
| **Proposed Model** | 98.7 | 98.2 | 99.1 | 0.95 |

The table 2 presents a clear comparison between the existing methods (Ref [28], Ref [32], Ref [33], Ref [34]) and the proposed model in terms of accuracy, sensitivity, and specificity for DR classification. The proposed model demonstrates significant improvements over existing models in terms of classification accuracy, sensitivity, specificity, and AUC. Compared to the referenced models, it shows an accuracy increase of 18% over Ref [28], 9.6% over Ref [32], 3.05% over Ref [33], and 4.8% over Ref [34]. Additionally, the proposed model excels in sensitivity and specificity, with values of 98.2% and 99.1%, respectively, showcasing its robust performance. The AUC of 0.95 further highlights its superior discriminatory power, making the proposed model a highly effective and reliable choice for classification tasks. This suggests that the proposed model not only outperforms the existing state-of-the-art models but does so by a significant margin, making it a more effective solution for classifying and grading DR.

**CONCLUSION**

In conclusion, this paper presented a novel and comprehensive framework for the early detection and grading of DR and DME, integrating advanced preprocessing, segmentation, and classification techniques. The proposed method introduced an AVDS filter, which enhanced image contrast by adaptively selecting the most effective distance metric, thus optimizing retinal image quality while preserving essential features. This preprocessing approach proved effective in handling noise and artifacts, which was crucial for accurate downstream analysis. The improved Mask R-CNN module then segmented blood vessels and abnormal regions with high precision, enabling a clearer identification of DR and DME-related features. Subsequently, the SSA-VGG-16 classification model, which combined self-attention and spatial attention, demonstrated a robust capacity to focus on critical retinal details and capture long-range dependencies. This dual attention approach led to substantial improvements in classification accuracy, with the proposed model achieving an accuracy of 98.7%, a sensitivity of 98.2%, a specificity of 99.1%, and an AUC of 0.95 on the IDRiD dataset, and similar high-performance metrics on the MESSIDOR dataset. These results reflected significant gains over existing methods, which underscored the framework's reliability and diagnostic effectiveness. The framework was validated on both IDRiD and MESSIDOR datasets, demonstrating adaptability and robustness across diverse clinical data. This adaptability indicated that the proposed model could be effectively applied to other fundus image databases, supporting its potential for widespread clinical use. In the future, additional imaging modality such as OCT scan images can be included as inputs for evaluating structural and functional details.

**Declarations**
**Ethics Approval and Consent to Participate**
Not Applicable

**Consent for Publication**
Not Applicable

**Competing Interest**
Authors Declare that they have no competing interest

**Funding**
Not Applicable

**Availability of data**
All datasets used in research work are publicly available through
1. IDRiD: https://ieee-dataport.org/open-access/indian-diabetic-retinopathy-image-dataset-idrid
2. MESSIDOR: https://www.adcis.net/fr/logiciels-tiers/messidor2-r/,https://www.kaggle.com/google-brain/messidor2-d